\documentclass[11pt,a4paper]{article}

\usepackage[utf8]{inputenc}   
\usepackage[T1]{fontenc}      
\usepackage{lmodern}          
\usepackage{amsmath,amssymb,amsthm} 
\usepackage{bm}               
\usepackage{graphicx}         
\usepackage{booktabs}         
\usepackage{hyperref}         
\usepackage{natbib}           
\usepackage{geometry}         
\title{Bayesian Network Propensity Score to Evaluate Treatment Effects in Observational Studies}

\author{
  Clelia Di Serio$^{*,\dagger}$,
  Federica Cugnata$^{\dagger}$,
  Pier Luigi Conti$^{\ddagger}$,
  Alberto Briganti$^{\S}$,\\
  Fulvia Mecatti$^{\P}$,
  Paola Vicard$^{\|}$,
  Paola Maria Vittoria Rancoita$^{\dagger}$
}

\date{}

\begin{document}
\maketitle
\begin{center}
\noindent $^{*}$ Corresponding author. E-mail: \texttt{clelia.diserio@unisr.it} \\[6pt]
$^{\dagger}$ Centre for Statistics in the Biomedical Sciences, Vita-Salute San Raffaele University, Milan, Italy \\[6pt]
$^{\ddagger}$ Sapienza University of Rome, Rome, Italy \\[6pt]
$^{\S}$ Vita-Salute San Raffaele University, Milan, Italy \\[6pt]
$^{\P}$ University of Milano-Bicocca, Milan, Italy \\[6pt]
$^{\|}$ Roma Tre University, Rome, Italy

\end{center}

\begin{abstract}
This paper focuses on the Bayesian Network Propensity Score (BNPS), a novel approach for estimating treatment effects in observational studies characterized by unknown (and likely unbalanced) designs and complex dependency structures among covariates. Traditional methods, such as logistic regression, often impose rigid parametric assumptions that may lead to misspecification errors, compromising causal inference. Recent classical and machine learning alternatives, such as  boosted CART,  
random forests, 
and Stable Balancing Weights,  
seem to be attractive in a  predictive perspective, but they typically lack asymptotic properties, such as consistency, efficiency, and valid variance estimation. In contrast, the recently proposed BNPS to estimate propensity scores uses Bayesian Networks to flexibly model conditional dependencies while preserving essential statistical properties such as consistency, asymptotic normality and asymptotic efficiency. 
Combined with the H{\'a}jek estimator, BNPS enables robust estimation of the Average Treatment Effect (ATE) in scenarios with strong covariate interactions and unknown data-generating mechanisms. Through extensive simulations across fifteen realistic scenarios and varying sample sizes, BNPS consistently outperforms benchmark methods in both empirical rejection rates and coverage accuracy. Finally, an application to a real-world dataset of 7,162 prostate cancer patients from San Raffaele Hospital (Milan, Italy) demonstrates BNPS’s practical value in assessing the impact of pelvic lymph node dissection on hospitalization duration and biochemical recurrence. The findings support BNPS as a statistically robust, interpretable and transparent alternative for causal inference in complex observational settings, enhancing the reliability of evidence from real-world biomedical data.

\end{abstract}

\section{Introduction}

Observational studies are increasingly relied upon in biomedical research when randomized controlled trials (RCTs) are unfeasible due to ethical, financial, or practical limitations. While they offer valuable clinical insights, the lack of randomization poses major methodological challenges for causal inference. In particular, when the data-generating process 
is unknown, conclusions may be compromised by unmeasured confounding, selection bias, or model misspecification.
To face these issues, propensity scores (PS) are widely used to adjust for differences in covariate distributions between treatment groups. However, PS are often estimated using logistic regression, a method based on strong parametric assumptions and a fixed functional form. This rigidity may fail to reflect the true underlying dependence structure among variables particularly in complex biomedical settings, thus limiting the reliability of treatment effect estimation. Recent developments in statistics and machine learning have introduced more flexible tools for PS estimation, including boosted CART, random forests, and stable balancing weights (SBW); cfr.  \cite{breiman01}, \cite{lee10}, \cite{west10}, \cite{rf},  \cite{zubi15}, \cite{sbw}. Although these methods improve predictive performance, they often behave as “black boxes”, lacking both interpretability and rigorous proofs of basic fundamental properties such as consistency, asymptotic normality and control over distributional properties, although these are of fundamental importance for a well principled causal inference.


In this paper, we consider a novel approach for estimating propensity scores using Bayesian Networks (BNs), namely BNPS (Bayesian Network Propensity Score). Such an approach has been recently proposed in \cite{noi25}, where its theoretical properties are studied. 
BNs offer a statistically coherent framework for modeling conditional dependencies among variables, preserving rigorous properties such as consistency and efficiency through maximum likelihood estimation. At the same time, they provide the flexibility of a data-driven tool, by adapting to the data structure without relying on predefined parametric assumptions. This allows the method to effectively capture the complex and often hierarchical relationships typical of biomedical data in a transparent, interpretable, and non-black-box manner. Moreover, the graphical nature of Bayesian Networks offers a transparent and interpretable representation of variable relationships. At the same time, the approach preserves distributional properties grounded in maximum likelihood estimation, providing a statistically rigorous foundation for valid and efficient causal inference.
In this contribution we highlight how, by combining BNPS and H{\'a}jek estimator,  an optimal setting is obtained, that enables  robust estimation of the Average Treatment Effect (ATE) in observational settings. To assess its performance, we will conduct a comprehensive simulation study comparing our method with several well-reputed alternatives:
$i)$ logistic regression PS with and without covariates in a weighted logistic regression (WLR); $ii)$ boosted CART (via the twang package \cite{twang}) with and without covariates in a WLR; $iii)$ random forests (via randomForest, cfr. \cite{rf}) with and without covariates in a WLR; $iv)$ stable balancing weights (via \cite{sbw}).

The simulation design is studied to reflect a wide range of realistic scenarios, by varying sample sizes, covariate interactions, and treatment effects.

Finally, we apply the proposed framework to a real-world dataset of 7,162 prostate cancer patients treated at San Raffaele Hospital (Milan, Italy), to evaluate the effect of pelvic lymph node dissection on hospitalization duration and biochemical recurrence. This application illustrates the utility of the method in clinical practice, where data complexity and confounding are common.

The  paper is organized as follows. Section 1 introduces the methodological framework for PS estimation using Bayesian Networks and H{\'a}jek-based ATE estimation. Section 2 presents the simulation study and comparative results. Section 3 reports the real-world data analysis in urology. Section 4 concludes with a discussion of implications, limitations, and future research directions.

\section{Bayesian Network to empower propensity score estimation}
In very recent literature \cite{noi25}, BNPS method has been proposed leveraging BNs for estimating propensity scores, which are crucial for causal inference in observational studies with binary outcomes and discrete covariates. As already said, traditional methods such as logistic regression are criticized for their rigid assumptions and inability to capture complex interactions involving covariates. BNs are 
a flexible statistical tool that can model conditional dependencies among covariates and treatment assignments without predefined functional forms. 
In the above mentioned paper, it is shown that the methodology based on BNs allows the use of the Maximum Likelihood Method, that leads to asymptotically normal and efficient estimators of propensity scores, even when the structure of relationships among covariates is not known {\em a priori}, but learned from observed data. 
This would potentially lead, in its turn, to more accurate estimates of the Average Treatment Effect (ATE). The theoretical advantages of BNs can be identified in avoiding misspecification errors, which are common in logistic models, and in making efficient the use of inverse probability weighting for ATE estimation, particularly through H{\'a}jek and Horvitz-Thompson estimators  based on estimated propensity scores. Empirical studies using simulations based on a real-world dataset of prostate cancer patients are conducted to compare the performance of BNs against logistic regression in estimating ATE. Results show that propensity scores estimated {\em via} BNs lead to more accurate and efficient ATE estimates compared to those derived from logistic models. The H{\'a}jek estimator, in particular, exhibits higher stability and robustness against model misspecification. Empirical coverage and rejection rates for treatment effect tests are also higher with the BN approach, indicating more reliable inference. The study highlights how BNs can better handle the intricacies of real-world data, thus offering a significant improvement over traditional methods.
In conclusion, the work in \cite{noi25} provides a sound theoretical foundation for the application of BNs in the estimation of propensity scores, stressing their ability to deliver universally consistent and asymptotically efficient estimates. BNs offer a flexible data-driven approach, which allows to more accurately reflect the underlying causal structure of the data, thereby reducing the risk of misspecification errors that can lead to biased estimates and inefficiencies in the causal inference process.
In addition to the above, it should be emphasized the critical importance of accurately capturing complex dependencies, to avoid the pitfalls associated with model misspecification. In observational studies, where randomized controlled conditions are absent, accounting for these dependencies is crucial to single out the causal effect of treatments. BNs, with their graphical representation of probabilistic relationships, inherently address these challenges by allowing for the inclusion of multiple covariate interactions and dependencies without pre-imposed constraints.
BNs significantly improve  the precision and reliability of Average Treatment Effect (ATE) estimation, as well as the power in testing  statistical hypotheses. This advancement not only enhances the validity of causal inferences drawn from observational studies but also provides a more robust framework for policy-making and clinical decision-making based on real-world evidence.
BNPS framework can be extended also to accommodate more complex scenarios, such as those involving multiple treatments and non-binary outcomes.

Despite the excellent theoretical properties shown in \cite{noi25}, a very important point still stands, namely a comparison of the BNPS approach with machine learning and other methods currently proposed in the literature. This is the main motivation of the present paper, namely providing a solid simulation study aiming at explaining why and when a methodology based on BNs behaves better than other methodologies for propensity score estimation. For the sake of completeness, in the present study we have included not only widely used machine learning methods, but also the traditional, and still now widely used, logistic regression.  

\section{Simulation study}
\label{subs:controlled-sim-plan}

To investigate the performance of different methodologies, a simulation study has been performed. In this study, several scenarios with different nature and strength of the dependence structure involving treatment,  covariates and outcome, have been considered. Among the scenarios, the simulation explores also complex (but realistic) cases where the treatment assignment is strongly affected by
covariates interaction, and the outcome may depend on the same covariates.

\subsection{Simulation study plan}
\label{subs:controlled-sim-plan-description}

\begin{figure}[!h]
\begin{center}
\includegraphics[width=0.5\textwidth]{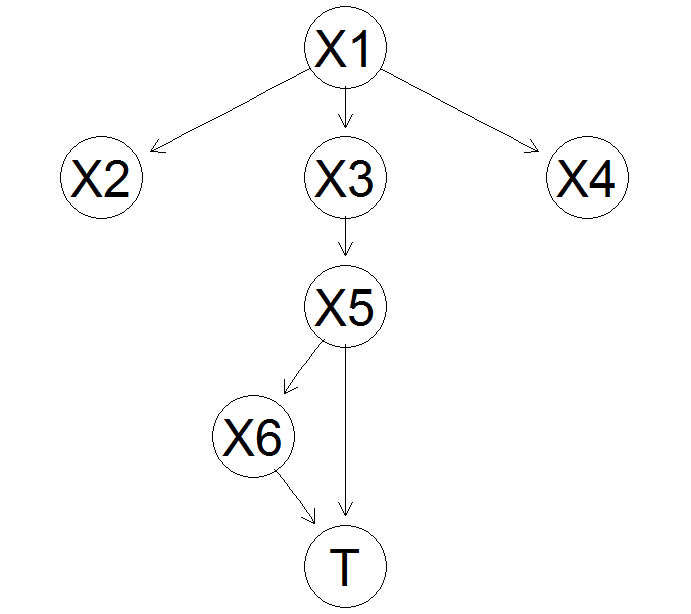}
\end{center}
\caption{DAG used for simulating the data.}
\label{fig:DAGsimulato}       
\end{figure}

In all simulation scenarios, we considered six categorical covariates ($X_i$ with $i=1,\ldots,6$) and a binary treatment assignment variable (T). Data were randomly generated from the Bayesian network shown in Figure \ref{fig:DAGsimulato}, with parameters of the conditional distributions reported in Table \ref{tab:condProbDAG}. 

In order to define the BN, a directed acyclic graph (DAG) with 7 nodes has been first generated, where the number of categories for each variable $X_i$, $i=1,\ldots,6$, is either 2 or 3. 
Let $pa(X_i)$ be the {\em parent set} of $X_i$, {\em i.e.} the set of all nodes (covariates) with a directed arc pointing to $X_i$. 
Then, the conditional probability distributions $\{P(X_i|\bm{X}_{pa(X_i)})\}_{i=1}^6$ has been sampled from a Dirichlet distribution with concentration parameters equal to 1.
The conditional probabilities of the treatment given the two covariates $X_5$ and $X_6$ (Figure \ref{fig:DAGsimulato}) have been defined to create a strong conditioning on the interaction between $X_5$ and $X_6$.

\begin{table}[!h]
\begin{center}
\caption{Parameters of the conditional distributions of the Bayesian network in Figure \ref{fig:DAGsimulato}.}
\label{tab:condProbDAG}
\begin{tabular}{lc}
\hline\noalign{\smallskip}
	Event &	Probability	\\
\hline
$	X_1=1	$	&	0.6553	\\
$	X_2=1|X_1=0	$	&	0.9961	\\
$	X_2=1|X_1=1	$	&	0.4384	\\
$	X_3=1|X_1=0	$	&	0.2186	\\
$	X_3=2|X_1=0	$	&	0.6238	\\
$	X_3=1|X_1=1	$	&	0.2102	\\
$	X_3=2|X_1=1	$	&	0.0556	\\
$	X_4=1|X_1=0	$	&	0.1600	\\
$	X_4=2|X_1=0	$	&	0.6778	\\
$	X_4=1|X_1=1	$	&	0.4151	\\
$	X_4=2|X_1=1	$	&	0.5150	\\
$	X_5=1|X_3=0	$	&	0.4536	\\
$	X_5=1|X_3=1	$	&	0.6636	\\
$	X_5=1|X_3=2	$	&	0.9726	\\
$	X_6=1|X_5=0	$	&	0.3967	\\
$	X_6=1|X_5=1	$	&	0.7000	\\
\hline
$	T=1|X_5=0,X_6=0	$	&	0.2	\\
$	T=1|X_5=0,X_6=1	$	&	0.6	\\
$	T=1|X_5=1,X_6=0	$	&	0.4	\\
$	T=1|X_5=1,X_6=1	$	&	0.2	\\
\hline
\end{tabular}
\end{center}
\end{table}

Once the treatment variable and the other covariates were generated, the binary potential outcomes $Y_{(0)}$ and $Y_{(1)}$, as specified below, 
possess Bernoulli distribution:
\begin{eqnarray}
Y_{(k)}|X_5, X_6 \sim B(P(Y_{(k)}|  X_5, X_6)), \quad k=0,1,
 \nonumber
\end{eqnarray}
which is linked to the covariates through the following logistic models:
\begin{eqnarray}
logit P(Y_{(0)}=1|X_5, X_6) = \alpha_0 +\beta_1 X _5 + \beta_2 X _6, \;\;\;  \label{eq:22}    \\
logit P(Y_{(1)}=1|X_5, X_6) = \alpha_0 +\alpha_1+\beta_1 X _5 + \beta_2 X _6. \label{eq:23}
\end{eqnarray}
 To derive the true ATE given the values of the coefficients of the models, the probabiloties $\theta_k=P(Y_{(k)}=1)$, $k=0,1$, have been computed by marginalizing $P(Y_{(k)}=1 | X_5, X_6)$ over $X_5, X_6$. 
 
 Fifteen scenarios (from S1 to S15) were defined by creating different combinations of: 1) 
 the subset of covariates (besides the treatment) influencing the outcome and the strength of this influence;  and 2) the strength of the treatment effect on the outcome. Namely, we defined scenarios were, except for the treatment, the potential outcomes depended on: (a) no variable among $X_1$, $\ldots$, $X_6$  (scenarios S1, S2, S3); (b) only the variable $X_5$ with moderate influence (scenarios S4, S5, S6); (c) only the variable $X_6$ with strong influence (scenarios S7, S8, S9); (d) two variables $X_5$ and $X_6$ with medium (scenarios S10, S11, S12) or strong influence (scenarios S13, S14, S15). 
For each of these forms of dependencies, three scenarios were defined in order to have either no, small or large treatment effect (i.e. ATE). The simulation parameters of the different scenarios are reported in Table \ref{tab:scenari}.
Finally, the observed outcome $Y$ is equal to $Y_{(0)}$ if $T=0$, and to $Y_{(1)}$ if $T=1$.

\begin{table}[!bth]
\caption{Parameter values of the models of $Y_{(0)}$ and $Y_{(1)}$  in the Equations (\ref{eq:22}) and (\ref{eq:23}) for simulating the fifteen scenarios.}
\label{tab:scenari}
\begin{center}
\begin{tabular}{cccccc}
\hline\noalign{\smallskip}
	&	$\alpha_0$	&	$\alpha_1$	&	$\beta_1$	&	$\beta_2$	&	True ATE\\
\hline
S1	&	-2	&	0	&	0	&	0	&	0.0000\\
S2	&	-2	&	1	&	0	&	0	&	0.1497\\
S3	&	-2	&	2.5	&	0	&	0	&	0.5033\\
\hline
S4	&	-2	&	0	&	1	&	0	&	0.0000\\
S5	&	-2	&	1	&	1	&	0	&	0.2009\\
S6	&	-2	&	2.5	&	1	&	0	&	0.5318\\
\hline
S7	&	-2	&	0	&	0	&	-2	&	0.0000\\
S8	&	-2	&	1	&	0	&	-2	&	0.0791\\
S9	&	-2	&	2.5	&	0	&	-2	&	0.3042\\
\hline
S10	&	-2	&	0	&	1	&	-2	&	0.0000\\
S11	&	-2	&	1	&	1	&	-2	&	0.1131\\
S12	&	-2	&	2.5	&	1	&	-2	&	0.3857\\
\hline
S13	&	-2	&	0	&	2	&	-3	&	0.0000\\
S14	&	-2	&	1	&	2	&	-3	&	0.1104\\
S15	&	-2	&	2.5	&	2	&	-3	&	0.3482\\
\hline
\end{tabular}
\end{center}
\end{table}

Five sample sizes ($n$ = 250, 500, 1000, 2500, 5000) were considered, with 1000 Monte Carlo runs for each combination of scenario and sample size.
In each simulated dataset, the treatment effect was evaluated through the following methods: 
\begin{itemize}
    \item our proposed methodology BNPS, {\em i.e.} by estimating the propensity score with the Bayesian Network learned using the Tabu search greedy algorithm  with the BIC score function (by using the \texttt{bnlearn} package, \cite{scutari2010}), and then applying the H\'{a}jek estimator (\texttt{BN + H\'{a}jek}); 
\item weighted logistic regression (WLR) (by using the \texttt{survey} package \cite{lumley2004}) with weights defined based on the estimated propensity score, and with or without including other covariates (besides treatment) in the logistic model. Three alternative standard methods for propensity score estimation have been considered: 1) logstic regression (\texttt{log + WLR}), 2) boosted CART implemented in the \texttt{twang} package \cite{twang}  (\texttt{boostCart  + WLR}), 3) random Forest {\em via} the \texttt{randomForest package} \cite{rf},  (\texttt{randomForest  + WLR});
\item Stable Balancing Weights approach, implemented in the \texttt{sbw} package \cite{sbw}, which directly optimizes covariate balance through weights computation, without relying on propensity score estimation (\texttt{SBW}).

For comparing the performances of all methods, we computed the empirical rejection rate (ERR) for each combination of scenario and sample size. The ERR was defined as the proportion of times that method rejected the hypothesis of no treatment effect over the 1000 Monte Carlo runs. Since the methods \texttt{BN + H\'{a}jek} and \texttt{SBW} also directly provide a confidence interval for the ATE, for these two methods we additionally computed the empirical coverage (EC). The EC was defined as the proportion of times the 95\% confidence interval of the ATE contained the true value of the ATE (of the corresponding scenario) over the 1000 Monte Carlo runs.
\end{itemize}

\subsection{Results and discussion}
\label{subs:controlled-results}
Tables \ref{tab:ERR1} and \ref{tab:ERR2} report the empirical rejection rate (ERR) of all methods computed in the fifteen scenarios. 

\begin{figure}[!h]
\begin{center}
\includegraphics[width=0.5\textwidth]{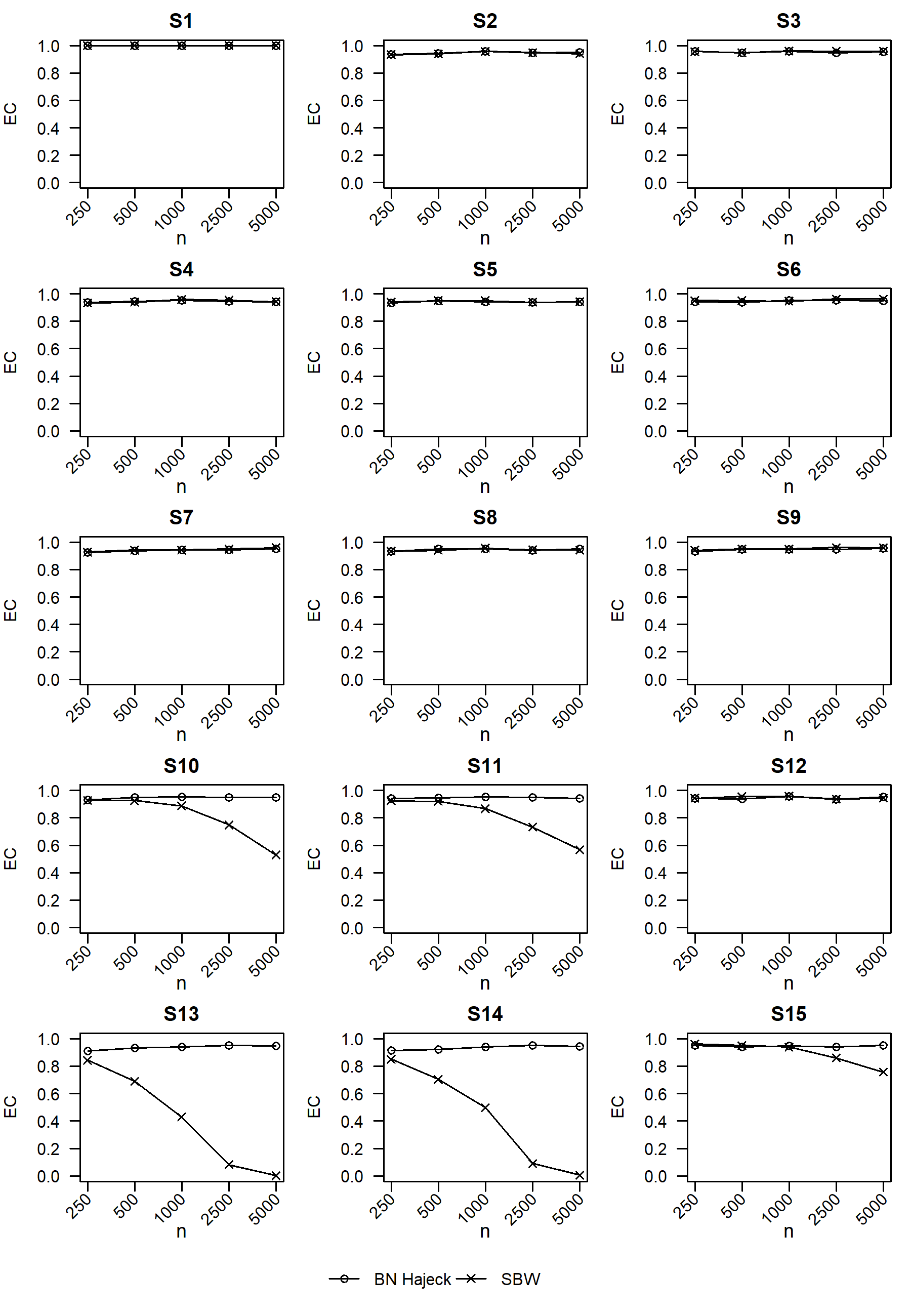}
\end{center}
\caption{Empirical coverage (EC) of the ATE across 15 simulated scenarios (S1–S15), evaluated at increasing sample sizes (n = 250, 500, 1000, 2500, 5000). Results are reported for two estimators: SBW and BN + H\'{a}jek}
\label{fig:EC}       
\end{figure}

In the first place, it should be noted that all methods, except \texttt{BN + H\'{a}jek} and \texttt{SBW}, tend to incorrectly identify a treatment effect in scenario S1, which is characterized by the absence of both treatment effect (ATE=0) and other covariates impacts on the outcome. In all other scenarios with no treatment effect (S4, S7, S10 and S13), only the \texttt{randomForest + WLR} generally possesses an ERR much higher that the nominal $5\%$, and, in some scenarios, the ERR even increases with the sample size. Overall, the methods \texttt{SBW} and \texttt{log + WLR noCov} (which does not use the covariates in the weighted logistic regression) exhibit a high ERR in scenarios S10 and S13, corresponding to a stronger  influence of the covariates on the outcome. Such a poor behavior of these methods becomes more and more evident as the sample size increases (since ERR increases instead of decreasing). As far as method \texttt{log + WLR} is concerned, this limitation is overcome by including the covariates in the weighted logistic regression (\texttt{log + WLR noCov}). In addition, in the same scenarions the EC for \texttt{SBW} decreases as the sample size increases (see Figure \ref{fig:EC}), this suggesting a bias in the estimation. 

A similar pattern in EC of \texttt{SBW} is also seen in scenarios S11 and S14, where the treatment effect is small and there is a strong influence of the covariates on the outcome (see Figure \ref{fig:EC}). In general, in all scenarios with small ATE, the \texttt{randomForest + WLR} had usually an ERR much lower than all other methods. In particular, its ERR correctly increases as the sample size does, but with a smaller rate in comparison to the others.

Overall, only  \texttt{BN + H\'{a}jek} is able to achieve a good performance in terms of both ERR and EC in all scenarios. Moreover, its performance always improves as the sample size increases, usually achieving acceptable results already for a sample size $n=250$,  or, in some complex scenarios (small ATE and some influence of other covariates on the outcome), for $n=500$.

\begin{table}[!h]
\centering
\caption{Empirical rejection rate (ERR) for scenarios S1-S9.\label{tab:ERR1}}
\scriptsize
\setlength\tabcolsep{4pt} 
\begin{tabular}{c c c c c c c c c c}
\toprule
 &  &  & \multicolumn{2}{c}{\textbf{log + WLR}} &
   \multicolumn{2}{c}{\textbf{boostCart + WLR}} &
   \multicolumn{2}{c}{\textbf{randomForest + WLR}} & \\
\cmidrule(lr){4-5}\cmidrule(lr){6-7}\cmidrule(lr){8-9}
Scenario & $n$ & \textbf{BN + H\'{a}jek}&
noCov & Cov &
noCov & Cov &
noCov & Cov &
\textbf{SBW}\\
\midrule
S1 & 250  & 0.000 & 0.934 & 0.950 & 0.814 & 0.832 & 0.997 & 1.000 & 0.000\\
S1   & 500  & 0.000 & 1.000 & 1.000 & 1.000 & 1.000 & 1.000 & 1.000 & 0.000\\
S1              & 1000& 0.000 & 1.000 & 1.000 & 1.000 & 1.000 & 1.000 & 1.000 & 0.000\\
  S1            & 2500& 0.000 & 1.000 & 1.000 & 1.000 & 1.000 & 1.000 & 1.000 & 0.000\\
    S1          & 5000& 0.000 & 1.000 & 1.000 & 1.000 & 1.000 & 1.000 & 1.000 & 0.000\\
\midrule
S2 & 250  & 0.693 & 0.756 & 0.772 & 0.745 & 0.745 & 0.382 & 0.538 & 0.741\\
  S2                 & 500  & 0.946 & 0.968 & 0.972 & 0.955 & 0.959 & 0.582 & 0.676 & 0.968\\
    S2               & 1000& 1.000 & 1.000 & 1.000 & 1.000 & 1.000 & 0.816 & 0.803 & 1.000\\
      S2             & 2500& 1.000 & 1.000 & 1.000 & 1.000 & 1.000 & 0.956 & 0.904 & 1.000\\
        S2           & 5000& 1.000 & 1.000 & 1.000 & 1.000 & 1.000 & 0.995 & 0.980 & 1.000\\
\midrule
S3 & 250  & 1.000 & 1.000 & 1.000 & 1.000 & 1.000 & 0.993 & 1.000 & 1.000\\
  S3                 & 500  & 1.000 & 1.000 & 1.000 & 1.000 & 1.000 & 1.000 & 1.000 & 1.000\\
    S3               & 1000& 1.000 & 1.000 & 1.000 & 1.000 & 1.000 & 1.000 & 1.000 & 1.000\\
      S3             & 2500& 1.000 & 1.000 & 1.000 & 1.000 & 1.000 & 1.000 & 1.000 & 1.000\\
        S3           & 5000& 1.000 & 1.000 & 1.000 & 1.000 & 1.000 & 1.000 & 1.000 & 1.000\\
\midrule
S4      & 250  & 0.063 & 0.049 & 0.061 & 0.052 & 0.066 & 0.113 & 0.149 & 0.066\\
  S4                 & 500  & 0.056 & 0.052 & 0.057 & 0.055 & 0.062 & 0.082 & 0.149 & 0.062\\
    S4               & 1000& 0.047 & 0.030 & 0.041 & 0.047 & 0.051 & 0.071 & 0.124 & 0.042\\
      S4             & 2500& 0.056 & 0.044 & 0.047 & 0.048 & 0.055 & 0.119 & 0.090 & 0.047\\
        S4           & 5000& 0.059 & 0.058 & 0.061 & 0.060 & 0.063 & 0.218 & 0.096 & 0.058\\
\midrule
S5 & 250  & 0.828 & 0.853 & 0.873 & 0.830 & 0.850 & 0.470 & 0.684 & 0.857\\
  S5                 & 500  & 0.989 & 0.997 & 0.997 & 0.989 & 0.989 & 0.710 & 0.781 & 0.995\\
    S5               & 1000& 1.000 & 1.000 & 1.000 & 1.000 & 1.000 & 0.916 & 0.886 & 1.000\\
      S5             & 2500& 1.000 & 1.000 & 1.000 & 1.000 & 1.000 & 0.995 & 0.933 & 1.000\\
        S5           & 5000& 1.000 & 1.000 & 1.000 & 1.000 & 1.000 & 1.000 & 0.991 & 1.000\\
\midrule
S6 & 250  & 1.000 & 1.000 & 1.000 & 1.000 & 1.000 & 0.990 & 1.000 & 1.000\\
   S6                & 500  & 1.000 & 1.000 & 1.000 & 1.000 & 1.000 & 1.000 & 0.999 & 1.000\\
     S6              & 1000& 1.000 & 1.000 & 1.000 & 1.000 & 1.000 & 1.000 & 1.000 & 1.000\\
       S6            & 2500& 1.000 & 1.000 & 1.000 & 1.000 & 1.000 & 1.000 & 1.000 & 1.000\\
         S6          & 5000& 1.000 & 1.000 & 1.000 & 1.000 & 1.000 & 1.000 & 1.000 & 1.000\\
\midrule
S7      & 250  & 0.076 & 0.049 & 0.059 & 0.049 & 0.069 & 0.244 & 0.198 & 0.072\\
   S7                & 500  & 0.063 & 0.059 & 0.058 & 0.053 & 0.062 & 0.185 & 0.197 & 0.058\\
     S7              & 1000& 0.055 & 0.050 & 0.052 & 0.040 & 0.049 & 0.159 & 0.178 & 0.058\\
       S7            & 2500& 0.055 & 0.052 & 0.052 & 0.051 & 0.062 & 0.323 & 0.159 & 0.051\\
         S7          & 5000& 0.049 & 0.066 & 0.047 & 0.042 & 0.048 & 0.693 & 0.139 & 0.040\\
\midrule
S8  & 250  & 0.417 & 0.548 & 0.534 & 0.486 & 0.515 & 0.265 & 0.418 & 0.462\\
   S8                & 500  & 0.708 & 0.820 & 0.792 & 0.749 & 0.770 & 0.447 & 0.498 & 0.738\\
     S8              & 1000& 0.954 & 0.990 & 0.982 & 0.966 & 0.974 & 0.741 & 0.667 & 0.975\\
       S8            & 2500& 1.000 & 1.000 & 1.000 & 1.000 & 1.000 & 0.996 & 0.807 & 1.000\\
         S8          & 5000& 1.000 & 1.000 & 1.000 & 1.000 & 1.000 & 1.000 & 0.899 & 1.000\\
\midrule
S9  & 250  & 1.000 & 1.000 & 1.000 & 1.000 & 1.000 & 0.944 & 0.996 & 1.000\\
S9                   & 500  & 1.000 & 1.000 & 1.000 & 1.000 & 1.000 & 0.997 & 0.999 & 1.000\\
  S9                 & 1000& 1.000 & 1.000 & 1.000 & 1.000 & 1.000 & 1.000 & 1.000 & 1.000\\
    S9               & 2500& 1.000 & 1.000 & 1.000 & 1.000 & 1.000 & 1.000 & 1.000 & 1.000\\
      S9             & 5000& 1.000 & 1.000 & 1.000 & 1.000 & 1.000 & 1.000 & 1.000 & 1.000\\
\bottomrule
\end{tabular}
\end{table}

\begin{table}[htbp]
\centering
\caption{Empirical rejection rate (ERR) for scenarios S10-S15.\label{tab:ERR2}}
\scriptsize
\setlength\tabcolsep{4pt} 
\begin{tabular}{c c c c c c c c c c}
\toprule
 &  &  & \multicolumn{2}{c}{\textbf{log + WLR}} &
   \multicolumn{2}{c}{\textbf{boostCart + WLR}} &
   \multicolumn{2}{c}{\textbf{randomForest + WLR}} & \\
\cmidrule(lr){4-5}\cmidrule(lr){6-7}\cmidrule(lr){8-9}
Scenario & $n$ & \textbf{BN + H\'{a}jek}&
noCov & Cov &
noCov & Cov &
noCov & Cov &
\textbf{SBW}\\
\midrule
S10 & 250  & 0.069 & 0.077 & 0.063 & 0.043 & 0.069 & 0.191 & 0.189 & 0.073 \\
S10 & 500  & 0.050 & 0.108 & 0.052 & 0.044 & 0.055 & 0.135 & 0.165 & 0.074 \\
S10 & 1000 & 0.046 & 0.191 & 0.049 & 0.042 & 0.051 & 0.142 & 0.145 & 0.113 \\
S10 & 2500 & 0.052 & 0.400 & 0.055 & 0.044 & 0.054 & 0.710 & 0.144 & 0.252 \\
S10 & 5000 & 0.052 & 0.687 & 0.057 & 0.052 & 0.054 & 0.987 & 0.122 & 0.470 \\
\midrule
S11 & 250  & 0.598 & 0.817 & 0.691 & 0.663 & 0.668 & 0.329 & 0.506 & 0.771 \\
S11 & 500  & 0.881 & 0.982 & 0.917 & 0.909 & 0.901 & 0.533 & 0.660 & 0.963 \\
S11 & 1000 & 0.993 & 1.000 & 0.999 & 0.998 & 0.998 & 0.868 & 0.809 & 1.000 \\
S11 & 2500 & 1.000 & 1.000 & 1.000 & 1.000 & 1.000 & 1.000 & 0.909 & 1.000 \\
S11 & 5000 & 1.000 & 1.000 & 1.000 & 1.000 & 1.000 & 1.000 & 0.958 & 1.000 \\
\midrule
S12 & 250  & 1.000 & 1.000 & 1.000 & 1.000 & 1.000 & 0.958 & 0.997 & 1.000 \\
S12 & 500  & 1.000 & 1.000 & 1.000 & 1.000 & 1.000 & 0.999 & 0.999 & 1.000 \\
S12 & 1000 & 1.000 & 1.000 & 1.000 & 1.000 & 1.000 & 1.000 & 1.000 & 1.000 \\
S12 & 2500 & 1.000 & 1.000 & 1.000 & 1.000 & 1.000 & 1.000 & 1.000 & 1.000 \\
S12 & 5000 & 1.000 & 1.000 & 1.000 & 1.000 & 1.000 & 1.000 & 1.000 & 1.000 \\
\midrule
S13 & 250  & 0.089 & 0.186 & 0.076 & 0.043 & 0.084 & 0.182 & 0.195 & 0.157 \\
S13 & 500  & 0.066 & 0.401 & 0.054 & 0.041 & 0.062 & 0.145 & 0.151 & 0.311 \\
S13 & 1000 & 0.061 & 0.685 & 0.051 & 0.033 & 0.056 & 0.251 & 0.150 & 0.572 \\
S13 & 2500 & 0.048 & 0.976 & 0.049 & 0.030 & 0.045 & 0.934 & 0.138 & 0.919 \\
S13 & 5000 & 0.051 & 0.999 & 0.049 & 0.031 & 0.047 & 0.999 & 0.133 & 0.998 \\
\midrule
S14 & 250  & 0.623 & 0.885 & 0.652 & 0.606 & 0.625 & 0.221 & 0.487 & 0.882 \\
S14 & 500  & 0.865 & 0.997 & 0.902 & 0.882 & 0.888 & 0.473 & 0.662 & 0.995 \\
S14 & 1000 & 0.995 & 1.000 & 0.999 & 0.997 & 0.998 & 0.857 & 0.775 & 1.000 \\
S14 & 2500 & 1.000 & 1.000 & 1.000 & 1.000 & 1.000 & 1.000 & 0.909 & 1.000 \\
S14 & 5000 & 1.000 & 1.000 & 1.000 & 1.000 & 1.000 & 1.000 & 0.958 & 1.000 \\
\midrule
S15 & 250  & 1.000 & 1.000 & 1.000 & 1.000 & 1.000 & 0.904 & 0.993 & 1.000 \\
S15 & 500  & 1.000 & 1.000 & 1.000 & 1.000 & 1.000 & 0.998 & 1.000 & 1.000 \\
S15 & 1000 & 1.000 & 1.000 & 1.000 & 1.000 & 1.000 & 1.000 & 1.000 & 1.000 \\
S15 & 2500 & 1.000 & 1.000 & 1.000 & 1.000 & 1.000 & 1.000 & 1.000 & 1.000 \\
S15 & 5000 & 1.000 & 1.000 & 1.000 & 1.000 & 1.000 & 1.000 & 1.000 & 1.000 \\
\bottomrule
\end{tabular}
\end{table}

\section{Application to real data}
The proposed approach was applied to real observational data of 7162 prostate cancer patients who underwent radical prostatectomy at San Raffaele Hospital (Milan, Italy). The primary goal is to evaluate the effect of removing pelvic lymph nodes during prostate cancer surgery (treatment variable) on the hospital stay duration, accounting for possible confounding effects. As a matter of fact, extensive pelvic lymph node dissection may lead to a prolonged hospitalization, due to possible post-surgery complications. The outcome variable was categorized, by defining a long hospital stay as the one with a duration greater than the third quartile of the variable (in the present case, 10 days). As a secondary outcome, the biochemical recurrence at 5 years was considered. This outcome should not differ between the two treatment groups, since each patient should have received the best treatment according to his characteristics. This outcome variable was available only for a subset of 2741 patients. In the analysis, six pre-treatment variables potentially influential on the decision about the administrated treatment were considered: age, Charlson Comorbidity Index (CCI), biopsy Gleason score (bxgg), clinical stage (clinstage), total PSA (tpsa) and neoadjuvant hormonal therapy (NEOadjHT). Since decision making in clinical practice is commonly based on categorical or categorized covariates, all the quantitative covariates have been categorized by clinically meaningful cut-offs (age: $<$55, between 55 and 65, between 65 and 75, $\geq$75; CCI: 0, 1, $\geq$2; tpsa: $<$4, between 4 and 10, $>$10; bxgg: $\leq$6, equal to 7, $\geq$8).

\begin{figure}[!h]
\centering
\includegraphics[width=0.5\linewidth]{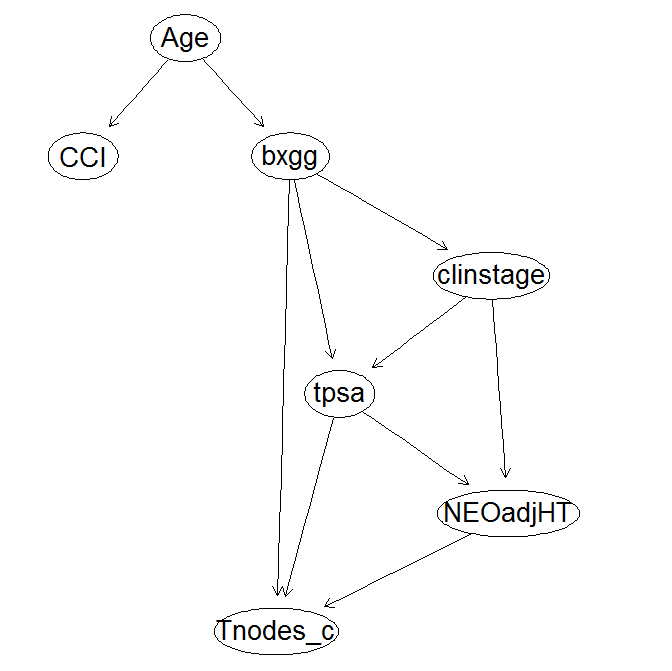}
\caption{Bayesian network obtained on the 7162 patients prostate cancer  through the TABU algorithm with the BIC score function. }\label{fig:BN_soloLos} 
\end{figure}

Figure \ref{fig:BN_soloLos} shows the Bayesian network for treatment variable and covariates, that was estimated through the Tabu greedy search (TABU) algorithm with BIC score function. It can be seen that a complex structure of interactions among treatments and pre-treatment covariates is present in this real setting. 
The estimated ATE corresponding to the outcome of long hospitalization is 
0.13 (95\% confidence interval = [0.03; 0.23]), confirming the expected effect of lymph node removal. 

\begin{figure}[h]
\centering
\includegraphics[width=0.5\linewidth]{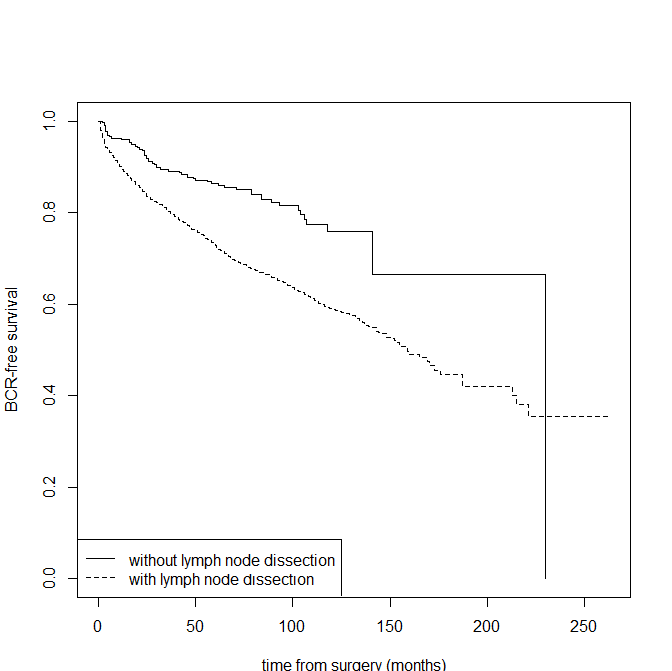}
\caption{Biochemical recurrence (BCR) free survival of the 2741 patients prostate cancer, for which this data was available. }\label{fig:BCR}
\end{figure}

As far as biochemical recurrence (BCR) is concerned, Figure \ref{fig:BCR} shows that patients who underwent lymph node dissection seem to have a worse outcome, if we do not account 
for possible confounding factors. The p-value of  the Fisher's exact test evaluating the simple association between the BCR at 5 years and the treatment is $<$0.0001, supporting the association between the two variables. 
On the other hand, when accounting for possible confounders, the estimated ATE related to 5-year BCR is 0.04 (95\% confidence interval = [-0.96; 1.03]), showing no treatment effect and thus supporting the hypothesis that each patient received the best treatment according to his characteristics.

\section{Discussion} This study introduces for the first time the Bayesian Network Propensity Score (BNPS) as an efficient, robust and flexible approach for Causal Inference in observational studies characterized by unbalanced designs and complex dependency structures. By modeling conditional dependence relationships among covariates without relying on rigid parametric assumptions, BNPS offers a statistically effective alternative to both traditional and machine learning–based methods for propensity score estimation.
Through a comprehensive simulation study involving fifteen realistic scenarios and different sample sizes, BNPS - combined with the H{\'a}jek estimator — consistently showed better performance in terms of empirical rejection rates and coverage probabilities. Notably, the method is characterized by a small, non-increasing bias in the estimation of average treatment effects. This clearly reflects the flexibility of BNPS, that avoids common misspecification errors in modelling propensity scores. The same does not occur, for instance, in using logistic regression or Stable Balancing Weights, because the corresponding bias may increase as the sample size does especially in the presence of model misspecification or complex covariates interactions.
The application to a real-world dataset of 7,162 prostate cancer patients further illustrates the practical value of BNPS in clinical research, where confounding, selection bias, and heterogeneous treatment pathways are often challenging. The method yielded stable and interpretable estimates for both short-term and long-term outcomes, reinforcing its potential for use in real-world health data analysis.
BNPS expands the set of statistical techniques available for causal inference, through the integration of interpretability, statistical rigor, and robustness to model misspecification. By controlling bias while maintaining consistency and efficiency, it supports clinicians and medical researchers in drawing robust causal conclusions from observational evidence. Future developments may explore extensions to continuous treatments, survival analysis, and longitudinal data frameworks.

\bibliographystyle{plainnat} 
\bibliography{references}    

\end{document}